\documentclass[fleqn,usenatbib]{mnras}

\usepackage{newtxtext,newtxmath}

\usepackage[T1]{fontenc}
\usepackage{ae,aecompl}

\usepackage{graphicx}	
\usepackage{amsmath}	
\usepackage{amssymb}	

\usepackage{threeparttable}

\title[Short title, max. 45 characters]{Local Stellar Kinematics and Oort Constants from the LAMOST A-type Stars}

\author[Wang et al.]{
F. Wang$^{1,2}$,
H.-W. Zhang$^{1,2}$\thanks{Corresponding authors: zhanghw@pku.edu.cn (HWZ); yanghuang @ynu.edu.cn (YH)},
Y. Huang$^{3}$\footnotemark[1],
B.-Q. Chen$^{3}$,
H.-F. Wang$^{3,4,5}$
and C. Wang$^{1,2,5}$
\\
$^{1}$Department of Astronomy,  School of Physics, Peking University, Beijing 100871, P.R. China.\\
$^{2}$Kavli Institute for Astronomy and Astrophysics,
Peking University,
Beijing 100871, P.R. China.\\
$^{3}$South-Western Institute for Astronomy Research,
Yunnan University,
Kunming, Yunnan 650091, P.R. China.\\
$^{4}$Department of Astronomy, China West Normal University, Nanchong 637009, P.R. China.\\
$^{5}$LAMOST Fellow
}

\date{Accepted 2021 March 16. Received 2021 March 15; in original form 2020 July 12}

\pubyear{2020}

\begin{document}

\maketitle
\begin{abstract}
We estimate the solar peculiar velocities and Oort constants using a sample of 5,627 A-type stars with $d<0.6\,\rm kpc$ and $|z|<0.1\,\rm kpc$, selected from the LAMOST surveys. The radial and tangential velocities of these A-type stars are fitted by using a non-axisymmetric model. The best-fitting result yields the solar peculiar velocities $(U_\odot,V_\odot,W_\odot)=(11.69\pm0.68, 10.16\pm0.51, 7.67\pm0.10)\,\rm km\,s^{-1}$ and Oort constants $A=16.31\pm0.89\,\rm km\,s^{-1}\,kpc^{-1}$, $B=-11.99\pm0.79\,\rm km\,s^{-1}\,kpc^{-1}$, $C=-3.10\pm0.48\,\rm km\,s^{-1}\,kpc^{-1}$, $K=-1.25\pm1.04\,\rm km\,s^{-1}\,kpc^{-1}$, respectively. $|K+C|>4\,\rm km\,s^{-1}\,kpc^{-1}$ means that there is a radial velocity gradient in the extended local disk, implying the local disk is in a non-asymmetric potential. Using the derived Oort constants, we derive the local angular velocity $\Omega\,{\approx}\,A-B=28.30\pm1.19\,\rm km\,s^{-1}\,kpc^{-1}$. By using A-type star sample of different volumes, we further try to evaluate the impacts of the ridge pattern in $R$-$V_{\phi}$ plane on constraining the solar motions and Oort constants. As the volume becomes larger toward the anti-center direction, the values of $A$ and $B$ become larger (implying a steeper slope of the local rotation curve) and the value of $V_\odot$ becomes smaller probably caused by the ridge structure and its signal increasing with distance.
\end{abstract}

\begin{keywords}
Galaxy: fundamental parameters --- Galaxy: kinematics and dynamics --- solar neighbourhood
\end{keywords}



\section{Introduction} \label{sec:intro}

The Galactic parameters are fundamental physical quantities in the kinematic and dynamic studies of the Milky Way (MW). The parameters include the distance from the Sun to the Galactic center $R_0$, the circular velocity at the location of the Sun $V_{\rm c}(R_0)$ and the solar peculiar velocities $(U_\odot,V_\odot,W_\odot)$ with respect to the local standard of rest (LSR). The LSR is defined as a reference system moving with a circular velocity $V_{\rm c}$ at $R_0$. One way to constrain $V_{\rm c}(R_0)$ is to measure the Oort constants (Oort 1927). The Oort constants, $A$ and $B$, describe the kinematic properties at the solar position of the MW. They can be used to constrain the circular velocity by combining the angular velocity $\Omega=A-B$ and $R_0$. Also different values of Oort constants reveal different kinematic nature of the disk stars of the MW, e.g. solid body rotation, Keplerian rotation or flat rotation curve.

\begin{table*}
    \centering
	\caption{Measurements of the solar motion in the literatures.}
	\label{tab:table_1}
	\begin{tabular}{ccccc} 
		\hline
		Source & Data & $U_\odot\,\rm(km\,s^{-1})$ & $V_\odot\,\rm(km\,s^{-1})$ & $W_\odot\,\rm(km\,s^{-1})$ \\
		\hline
        Mihalas \& Binney (1981) & Galactic Astronomy (2nd Ed.) & 9.00 & 12.00 & 7.0 \\
        Binney et al. (1997) & Main sequence stars & 11.00$\pm$0.60 & 5.30$\pm$1.70 & 7.00$\pm$0.60 \\
        Dehnen \& Binney (1998) & Main sequence stars & 10.00$\pm$0.60 & 5.25$\pm$0.62 & 7.17$\pm$0.38 \\
        Mignard (2000) & K0-K5 & 9.88 & 14.19 & 7.76 \\
        Piskunov et al. (2006) & Open clusters & 9.44$\pm$1.14 & 11.90$\pm$0.72 & 7.20$\pm$0.42 \\
        Bobylev \& Bajkova (2007) & F \& G dwarfs & 8.70$\pm$0.50 & 6.20$\pm$2.22 & 7.20$\pm$0.80 \\
        Francis \& Anderson (2009) & $Hipparcos$ & 7.50$\pm$1.00 & 13.50$\pm$0.30 & 6.80$\pm$0.10 \\
        Reid et al. (2009) & Masers & 9.0 & 20 & 10 \\
        Sch{\"o}nrich et al. (2010) & Main sequence stars & $11.10_{-0.75}^{+0.69}$ & $12.24_{-0.47}^{+0.47}$ & $7.25_{-0.36}^{+0.37}$ \\
        Breddels et al. (2010) & RAVE DR2 & 12.00$\pm$0.60 & 20.40$\pm$0.50 & 7.80$\pm$0.30 \\
        Bobylev \& Bajkova (2010) & Masers & 5.50$\pm$2.2 &11.00$\pm$1.70 & 8.50$\pm$1.20 \\
        Co\c{s}kuno\v{g}lu et al. (2011) & FGK dwarfs & 8.50$\pm$0.29 & 13.38$\pm$0.43 & 6.49$\pm$0.26 \\
        Bobylev \& Bajkova (2014) & Young objects & 6.00$\pm$0.50 & 10.60$\pm$0.80 & 6.50$\pm$0.30 \\
        Huang et al. (2015) & FGK dwarfs & 7.01$\pm$0.20 & 10.13$\pm$0.12 & 4.95$\pm$0.09 \\
        Bobylev (2017) & Cepheids & 7.90$\pm$0.65 & 11.73$\pm$0.77 & 7.39$\pm$0.62 \\
        Bobylev \& Bajkova (2019) & Open star clusters & 8.53$\pm$0.38 & 11.22$\pm$0.46 & 7.83$\pm$0.32 \\
        This study (2020) & A-type stars & 11.69$\pm$0.68 & 10.16$\pm$0.51 & 7.67$\pm$0.10 \\
		\hline
	\end{tabular}
\end{table*}

The solar peculiar velocities $(U_\odot,V_\odot,W_\odot)$ are important when converting the observed heliocentric velocities to the Galactic reference system. The radial and vertical velocities, $U_\odot$ and $W_\odot$, can be estimated directly by the mean values (e.g., by Gaussian fitting) of corresponding velocity distribution of stars in the solar neighborhood. But the component in the direction of Galactic rotation $V_\odot$ is more difficult to be determined, due to the asymmetric drift, the mean lag with respect to the LSR. This term is related to the velocity dispersion of the concerned stellar population. The slower rotation of stars will lead bias in estimating the value of $V_{\odot}$. Table\,1 summarizes recent results with different samples by different work (see similar tables in Francis \& Anderson 2009, Co\c{s}kuno\v{g}lu et al. 2011 and Huang et al. 2015). There are some differences of the results via various methods or tracers. One method to solve the asymmetric drift problem is to use the young stellar population and their velocity lag is small enough to be neglected. Bobylev (2017) and Bobylev \& Bajkova (2017, 2019) have adopted young tracers, e.g. Cepheids, OB stars and open star clusters, to constrain the solar motion and angular velocity. But the sizes of their samples are very limited (typically several hundred). 

Oort constants can be derived by the radial velocities and proper motions of stars in the solar vicinity. Oort (1927) first obtained the Oort constants, $A$ and $B$, and suggested a nearly flat rotation curve for the MW. After that, the most classical measurements of the constants are $A=14.82\pm0.84\,\rm km\,s^{-1}\,kpc^{-1}$ and $B=-12.37\pm0.64\,\rm km\,s^{-1}\,kpc^{-1}$ derived from the Cepheids with $Hipparcos$ proper motions (Feast \& Whitelock 1997). More recently, using the main-sequence stars with proper motions in Gaia DR1 Tycho-Gaia Astrometric Solution (TAGS), Bovy (2017) obtained more precise results: $A=15.3\pm0.4\,\rm km\,s^{-1}\,kpc^{-1}$, $B=-11.9\pm0.4\,\rm km\,s^{-1}\,kpc^{-1}$, $C=-3.2\pm0.4\,\rm km\,s^{-1}\,kpc^{-1}$ and $K=-3.3\pm0.6\,\rm km\,s^{-1}\,kpc^{-1}$. Similarly, Li et al. (2019) obtained $A=15.1\pm0.1\,\rm km\,s^{-1}\,kpc^{-1}$, $B=-13.4\pm0.1\,\rm km\,s^{-1}\,kpc^{-1}$, $C=-2.7\pm0.1\,\rm km\,s^{-1}\,kpc^{-1}$ and $K=-1.7\pm0.2\,\rm km\,s^{-1}\,kpc^{-1}$ by using a larger main-sequence star sample with proper motions from the Gaia DR2. The non-zero Oort constants $C$ and $K$ indicate the local Galactic disk is non-axisymmetric.

Gaia DR2 (Gaia Collaboration et al. 2018) provides position, parallax and proper motions for over a billion stars with unprecedented precision. The Large Sky Area Multi-Object Fiber Spectroscopic Telescope (LAMOST; Cui et al. 2012; Deng et al. 2012; Zhao et al. 2012; Liu et al. 2014) provides a large catalogue of over five million stars with radial velocities and atmospheric parameters (i.e. effective temperature $T_{\rm eff}$, surface gravity log\,$g$ and metallicity [Fe/H]) determined from low/medium resolution spectra. In this paper, a large sample of A-type stars is selected from the LAMOST surveys based on their stellar atmospheric parameters.
The A-type stars are relatively young and thus their asymmetric drift effects can be neglected in the current analysis. Combining the LAMOST and Gaia DR2, we can derive 6D information (positions, parallax, radial velocity and proper motions) of stars in the solar neighborhood to constrain the solar peculiar velocities and Oort constants.

This paper is organized as follows. The data employed in this work is described in Section 2. Our model and fitting results are shown in Sections 3 and 4, respectively. In Section 5, we discuss the impacts of the ridge structure on the determinations of the solar motions and Oort constants. Finally a summary is presented in Section 6.

\section{Data} \label{sec:style}

\subsection{LAMOST and Gaia data}
In this work, the data set we use is from the LAMOST DR4 value-added catalog (Xiang et al. 2017). The LAMOST is a 4\,m Schmidt telescope with a wide field ($5^{\circ}$ in diameter) at Xinglong Observing Station. The unique design of LAMOST enables it to take 4000 spectra in a single exposure at a resolving power $R{\sim}1800$ of wavelength coverage 3700-9000\,\AA. The stellar atmospheric parameters and radial velocities are derived from the LAMOST spectra using the LAMOST Stellar Parameter Pipeline at Peking University (LSP3; Xiang et al. 2015, 2017).

\begin{figure}
	\centering
	\includegraphics[width = \linewidth]{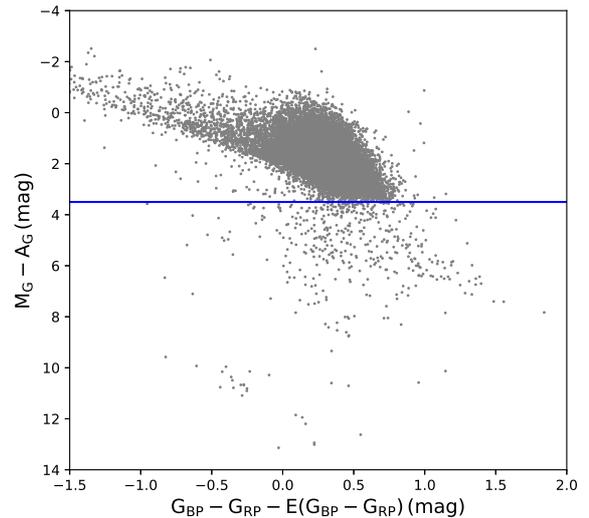}
	\caption{The color-magnitude diagram of our A-type star sample with the relative parallax error $\leq 10\%$ and parallax $>1\,\rm mas$. The gray dots are our sample stars and the blue line is determined empirically to exclude the contaminations from sub-dwarfs and white dwarfs.}
	\label{fig:1}
\end{figure}

The second data of Gaia mission was released on 25 April 2018 and provides a large sample of approximately 1.7 billion sources with precise position and apparent magnitude in $G$ and of over 1.3 billion sources with measurements of parallax and proper motions (Gaia Collaboration et al. 2018).  Using LAMOST and Gaia DR2 data, we can derive all components of velocities and distances for stars.

\begin{figure}
	\centering
	\includegraphics[width = \linewidth]{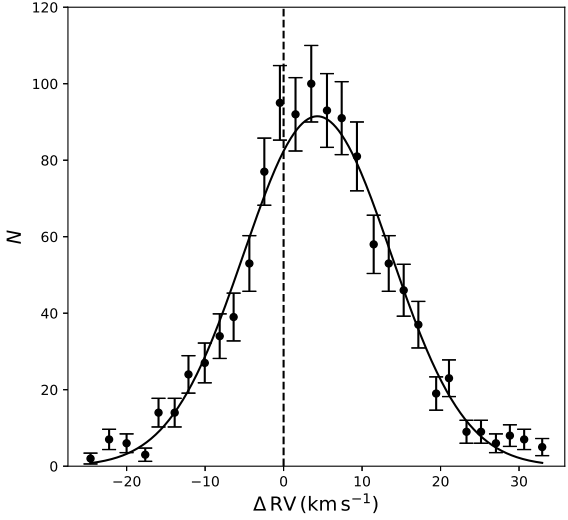}
	\caption{Distribution of the values of $\rm {\Delta}RV=RV_{APOGEE}-RV_{LAMOST}$ for 1,113 common stars after 3$\sigma$ clipping procedure. The dots represent the numbers of our sample in each velocity bin and the solid line shows the best Gaussian fitting result. The dashed line represents the zero value. }
	\label{fig:2}
\end{figure}

\subsection{Data selection}

The A-type star sample is selected by cuts in the effective temperature estimated by LSP3: i.e. $7200 \leq T_{\rm eff} \leq 9000$\,K and with spectral signal-to-noise ratio (SNR) $>30$ from the LAMOST surveys. The cuts result in 175,057 A-type star candidates. 
Then we cross match this catalogue with the Gaia DR2. We select the data with reliable parallax, i.e. the relative error $\leq 10\%$, and correct the parallax zero-point by adding an offset of 0.054 mas (Sch{\"o}nrich et al. 2019). We choose the data with $\rm 1/parallax<1\,kpc$ and the distance is directly estimated by the inverse of the parallax measurements. The distance yielded by this method is examined by the sample from Schonrich et al. (2019). For most (>90\%) stars, the differences between  those given one over parallax (correcting for zero-point offset) and those estimated by Schonrich et al. (2019) are smaller than 0.5\% for stars within 1\,kpc. In this degree, the distance adopted here should not induce any significant influence on the kinematic results.
Finally, we investigate the fraction of close binary systems (that would cause significant line-of-sight velocity variations and may alter our main results), using stars with multiple observations.  
Here, stars with the maximum differences of line-of-sight velocities larger than 20\,km\,s$^{-1}$ ($\sim$3$\sigma$ errors of line-of-sight velocity measurements) are simply assumed as close binary systems and the fraction is found only $6$\%.
The effect of close binary systems amongst our sample is therefore negligible.

For A-type star, the uncertainties of the atmospheric parameters derived by LSP3 (Xiang et al. 2015, 2017) are quite large (especially for $\log g$), we therefore try to exclude the sub-dwarfs and white dwarfs in the color-absolute magnitude diagram as shown in Fig.\,1 rather than in the $T_{\rm eff}-\log g$ diagram. The absolute magnitude is calculated by using the distance (from parallax). Then we apply a Rayleigh–Jeans colour excess (RJCE) method (Majewski et al. 2011) to estimate the extinction $A_{\rm G}$ and reddening $E(G_{\rm BP}-G_{\rm RP})$ using colour $(H - W2)$ from the Two Micron All Sky Survey (2MASS; Skrutskie et al. 2006) and the Wide-Field Infrared Survey Explorer (WISE; Wright et al. 2010). The intrinsic $(H - W_2)$ colour of A-type stars is first estimated by those A-type stars located at high Galactic latitude region (i.e., $|b| > 45 ^{\circ}$). Thus the reddening of those stars can be properly corrected by the values from the dust map of Schlegel et al. (1998). The extinction of the remaining stars are estimated by comparing their $(H - W2)$ colours to the intrinsic one. We discard the sample stars without $H$ or $W2$ magnitudes and the absolute magnitude larger than $3.5\,\rm mag$. Finally, 35,888 stars are left. After that, we further consider the selection criterion from Arenou et al. (2018) for the Gaia DR2 data as follows,
\begin{gather}
    \chi^2=\mathbf{astrometric\_chi2\_al}, \nonumber\\
    \nu=\mathbf{astrometric\_n\_good\_obs\_al}-5, \nonumber\\
    u=\sqrt{\chi^2/\nu}, \nonumber\\
    u<1.2\times\max(1,\exp{(-0.2(G-19.5))}), \nonumber\\
    \mathbf{visibility\_periods\_used}>8,
    \label{eq:Gaia}
\end{gather}
to filter bad solutions and contamination from double stars, binary stars and calibration problems. 

Then we try to examine the zero-point offset of the radial velocities $V_r$ estimated by LSP3 from the LAMOST surveys. Huang et al. (2018) have adopted nearly one hundred  radial velocity standard stars to examine the radial velocity zero point (RVZP) of the APOGEE survey. The examination shows the RVZP of APOGEE is very small. We thus adopt the radial velocity of APOGEE as the standard scale. We cross match the A-type sample with the APOGEE DR14 (Majewski et al. 2017; Abolfathi et al. 2018). The total number of the common stars is 1,185. We fit the differences of radial velocities between APOGEE and LAMOST with a Gaussian function by 3$\sigma$ clipping procedure. The distribution of the $V_r$ difference and the fitting result are shown in Fig.\,2. We find that the radial velocities of LAMOST are smaller than those of APOGEE by $4.30 \pm 0.33$\,km\,s$^{-1}$. So we add $4.30\,\rm km\,s^{-1}$ to the radial velocities of our sample as a correction. This correction value is similar to the result found by Huang et al. (2018).

We convert the right ascension R.A., declination Dec. and the corresponding proper motions to the Galactic longitude $l$, latitude $b$ and corresponding proper motions, $\mu_l$ and $\mu_b$, using galpy (Bovy 2015). The tangential velocities are calculated by $V_l=4.74{\mu_l}\cos{b}{\cdot}d$ and $V_b=4.74{\mu_b}{\cdot}d$ where $d$ is a heliocentric distance of the star in kpc. For the distance to the Galactic center $R_0$, we adopt $R_0=8.178\,\rm kpc$ (Gravity Collaboration et al. 2019). We also adopt the vertical position of the Sun $z_0=25\,\rm pc$ (Juri{\'c} et al. 2008). After imposing the requirements that the vertical distance $|z|<0.1\,\rm kpc$, the uncertainties of the vertical velocities $\epsilon_{V_l}$ and $\epsilon_{V_b}<5\,\rm km\,s^{-1}$ and the uncertainties of the radial velocities $\epsilon_{V_r}<10\,\rm km\,s^{-1}$, 19,034 stars are left. We further exclude 5 outliers with very large total velocities, $V_{\rm tot}=\sqrt{V_{r}^2+V_l^2+V_b^2}>150\,\rm km\,s^{-1}$. Finally, our sample has 19,029 A-type stars.

We note the current method (see Section 3) is only reasonable in the solar neighborhood ($d{\ll}R_0$) since the model considers the first-order Taylor series expansion of the velocity with respect to the solar position. On the other hand, a large volume is required to include enough sample stars to pin down the random errors of the estimated parameters. To balance the two facts, we decide to fit the observation data by the current model for 5,627 sample stars within 0.6\,kpc.

\section{Method} 

\label{sec:methods} 
\begin{figure*}
	\centering
	\includegraphics[width = \linewidth]{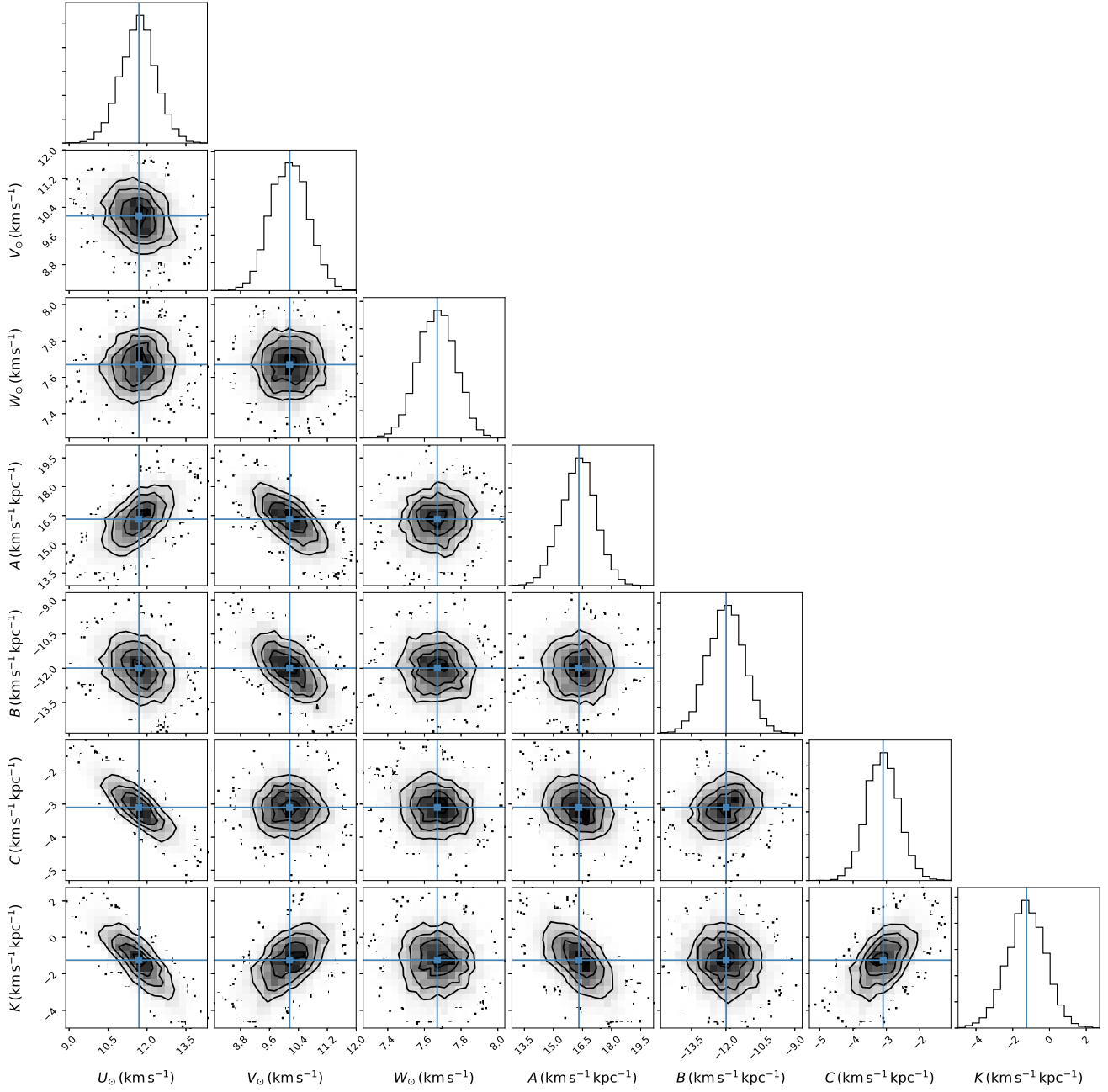}
	\caption{The MCMC results of the solar peculiar velocities $(U_\odot,V_\odot,W_\odot)$ and Oort constants $(A,B,C,K)$. The blue lines represent the median values we adopted for each parameter. The top panel of each column shows the probability distribution of the each parameter. The density distribution is the density of probability for every two parameters. The three black contours in each panel delineate the 1-$\sigma$, 2-$\sigma$ and 3-$\sigma$ confidence levels, separately.}
	\centering
	\label{fig:3}
\end{figure*}

\begin{figure*}
	\centering
	\includegraphics[width = \linewidth]{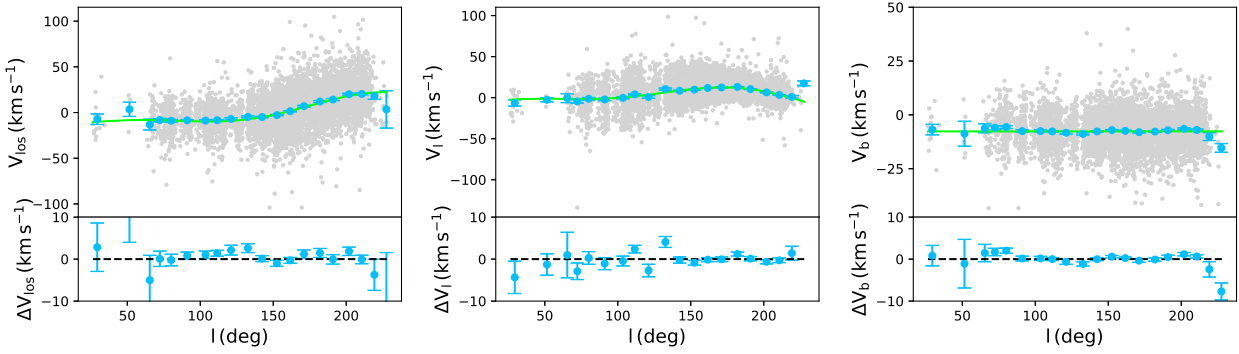}
	\caption{The $V_r$, $V_l$ and $V_b$ as a function of Galactic longitude $l$ from left to right panels. Bottom panels show the residual of velocity $\Delta=V_{\rm data}-V_{\rm model}$. The grey dots represent the velocities from our samples. The green lines show results from our best fitting model. The blue dots in the top panels represent the mean and uncertainty of velocities in each Galactic longitude bin and in the bottom represent the residual between data and model.  The black dashed lines represent the zero constant velocities. }
	\centering
	\label{fig:4}
\end{figure*}

Oort (1927) introduces the two Oort constants $A$ and $B$ to describe the rotation property of the MW disk under an axisymmetric assumption. Chandrasekhar (1942) generalizes Oort's analysis to a non-axisymmetric model and redefines the Oort constants, $A$ and $B$, and introduces two additional constants, $C$ and $K$. These constants are defined as:
\begin{gather}
    A=\frac{1}{2}\left(\frac{V_\phi}{R}-\frac{{\partial}V_\phi}{{\partial}R}-\frac{1}{R}\frac{{\partial}V_R}{{\partial}\phi}\right),  \\
    B=\frac{1}{2}\left(-\frac{V_\phi}{R}-\frac{{\partial}V_\phi}{{\partial}R}+\frac{1}{R}\frac{{\partial}V_R}{{\partial}\phi}\right),  \\
    C=\frac{1}{2}\left(-\frac{V_R}{R}+\frac{{\partial}V_R}{{\partial}R}-\frac{1}{R}\frac{{\partial}V_\phi}{{\partial}\phi}\right),  \\
    K=\frac{1}{2}\left(\frac{V_R}{R}+\frac{{\partial}V_R}{{\partial}R}+\frac{1}{R}\frac{{\partial}V_\phi}{{\partial}\phi}\right),
\end{gather}
where $V_R$ and $V_\phi$ are velocities with respect to the Galactocentric cylindrical coordinate system $R$ and $\phi$, respectively. If we consider $A$, $B$, $C$ and $K$ in an axisymmetric model that simply means $V_R=0$ and $V_\phi=V_\phi(R)$, and thus one can obtain $K=C=0$ and the expressions for the Oort constants $A$ and $B$ become 
\begin{gather}
A=\frac{1}{2}\left(\frac{V_\phi}{R}-\frac{{\partial}V_\phi}{{\partial}R}\right), B=-\frac{1}{2}\left(\frac{V_\phi}{R}+\frac{{\partial}V_\phi}{{\partial}R}\right),
\end{gather}
which are in line with the form of Oort (1927). In the solar vicinity, the three direction motions $V_r$, $V_l$ and $V_b$, can be represented by the solar peculiar velocities and Oort constants $A$, $B$, $C$ and $K$ as:
\begin{flalign}
V_{r} =&-U_\odot\cos{l}\cos{b}-V_\odot\sin{l}\cos{b}-W_\odot\sin{b} \nonumber \\
            &+d{\cdot}(K+A\sin{2l}+C\cos{2l})\cos{b}\cos{b},    \\
V_{l} =&U_\odot\sin{l}-V_\odot\cos{l} \nonumber \\
        &+d{\cdot}(A\cos{2l}+B-C\sin{2l})\cos{b},     \\
V_{b} =&U_\odot\cos{l}\sin{b}+V_\odot\sin{l}\sin{b}-W_\odot\cos{b}  \nonumber \\
         &-d{\cdot}(K+A\sin{2l}+C\cos{2l})\sin{b}\cos{b},
\end{flalign}
where $(U_\odot,V_\odot,W_\odot)$ is the solar velocity respect to the LSR in the Cartesian coordinate system. $V_\odot$ is in the direction of Galactic rotation, $W_\odot$ is positive toward the north Galactic pole and $U_\odot$ is positive toward the Galactic center.

\section{Results}

We fit the A-type star sample within $0.6\,\rm kpc$ using the model (Eqs. 7-9) described in Section 3. We adopt a maximum likelihood method to fit the model based on observational velocities $(V_r, V_l, V_b)$ and positions $(l, b, d)$ and obtain the best-fitting values of the solar peculiar velocities and Oort constants. The likelihood is written to a logarithm form to make calculation efficiently. The log likelihood function $\mathcal{L}$ is defined as follow,
\begin{flalign}
    \ln{\mathcal{L}} &=-\frac{1}{2}\sum_i{\Bigg(
\frac{(V_{r,{\rm obs},i}-V_{r,{\rm model},i})^2}{\sigma^2_{V_{r},{\rm tot},i}}+\ln{\big(2{\pi}{\sigma_{V_{r},{\rm tot},i}^2}\big)} }\nonumber\\
&{ +\frac{(V_{l,{\rm obs},i}-V_{l,{\rm model},i})^2}{\sigma^2_{V_l,{\rm tot},i}}
    +\ln{\big(2{\pi}{\sigma_{V_l,{\rm tot},i}^2}\big)} }\nonumber \\
    &{ +\frac{(V_{b,{\rm obs},i}-V_{b,{\rm model},i})^2}{\sigma^2_{V_b,{\rm tot},i}}
    +\ln{\big(2{\pi}{\sigma_{V_b,{\rm tot},i}^2}\big)} 
    \Bigg)},
\end{flalign}
where the subscript $\rm obs$ and $\rm model$ on the $V_{r}$, $V_l$ and $V_b$, respectively, mean the velocity from the observation data and from the model. The subscript $i$ is the $i$th star. In our fitting method, we also consider the velocity dispersion $\sigma_{V_R}$, $\sigma_{V_\phi}$ and $\sigma_{V_z}$ in the Galactocentric cylindrical coordinate system as free parameters to fit. We assume that they are simply constants at different Galactocentric distances and then transfer them into velocity dispersion components of $V_r$, $V_l$ and $V_b$. The $\sigma_{\rm tot}$ for each velocity component can be calculated by $\sigma_{\rm tot}^2=\sigma_{\rm model}^2+\epsilon_{\rm obs}^2$ where $\epsilon_{\rm obs}$ is the uncertainty of the data and $\sigma_{\rm model}$ is the velocity dispersion for $V_{r}$, $V_l$ and $V_b$ which can be calculated by the $\sigma_{V_R}$, $\sigma_{V_\phi}$ and $\sigma_{V_z}$ through coordinate transformations. The uncertainties of the free parameters ($A, B, C, K, U_\odot,V_\odot,W_\odot, \sigma_{V_R}, \sigma_{V_\phi}, \sigma_{V_z}$) are obtained via a Markov Chain Monte Carlo (MCMC) method.

The best-fitting result yields the solar peculiar velocity $(U_\odot,V_\odot,W_\odot)=(11.69\pm0.68, 10.16\pm0.51, 7.67\pm0.10)\,\rm km\,s^{-1}$. The results of Oort constants are $A=16.31\pm0.89\,\rm km\,s^{-1}\,kpc^{-1}$, $B=-11.99\pm0.79\,\rm km\,s^{-1}\,kpc^{-1}$, $C=-3.10\pm0.48\,\rm km\,s^{-1}\,kpc^{-1}$, $K=-1.25\pm1.04\,\rm km\,s^{-1}\,kpc^{-1}$. Fig.\,3 shows the MCMC results and the correlations between each two parameters. Some pair parameters show significant positive/negative correlations (e.g., $C$ and $K$, $U_\odot$ and $C$) and these correlations are probably due to incomplete sky coverage of the LAMOST which does not cover the Galactic centre direction at the low latitude regions. The best fitting model compared to the data is shown in Fig.\,4. We find that our model agrees with the data quite well.

\begin{table*}
	\centering
	\caption{The results for samples in different distance volumes.}
	\label{tab:table_2}
	\begin{tabular}{lcccccccc} 
		\hline
		 Range & Number & $U_\odot$ & $V_\odot$ & $W_\odot$ & $A$ & $B$ & $C$ & $K$\\
		 $\,\rm (kpc)$ & & $(\rm km\,s^{-1})$ & $(\rm km\,s^{-1})$ & $(\rm km\,s^{-1})$ & $(\rm km\,s^{-1}\,kpc^{-1})$ & $(\rm km\,s^{-1}\,kpc^{-1})$ & $(\rm km\,s^{-1}\,kpc^{-1})$ & $(\rm km\,s^{-1}\,kpc^{-1})$ \\
		\hline
         0-0.6 & 5627 & 11.69$\pm$0.68 & 10.16$\pm$0.51 & 7.67$\pm$0.10 & 16.31$\pm$0.89 & $-$11.99$\pm$0.79 & $-$3.10$\pm$0.48 & $-$1.25$\pm$1.04  \\
         0-0.7 & 8393 & 12.05$\pm$0.57 & 9.37$\pm$0.39 & 7.43$\pm$0.09 & 17.25$\pm$0.65 & $-$10.93$\pm$0.57 & $-$3.44$\pm$0.40 & $-$3.16$\pm$0.72  \\
         0-0.8 & 11658 & 11.84$\pm$0.48 & 8.74$\pm$0.34 & 7.24$\pm$0.07 & 18.53$\pm$0.44 & $-$9.95$\pm$0.43 & $-$3.08$\pm$0.36 & $-$3.53$\pm$0.54  \\
         0-0.9 & 15397 & 11.04$\pm$0.42 & 8.42$\pm$0.29 & 7.03$\pm$0.06 & 18.42$\pm$0.34 & $-$8.61$\pm$0.32 & $-$2.24$\pm$0.29 & $-$3.21$\pm$0.42  \\
         0-1.0 & 19029 & 9.79$\pm$0.37 & 8.50$\pm$0.26 & 6.95$\pm$0.05 & 17.94$\pm$0.30 & 
         $-$8.08$\pm$0.27 & $-$1.18$\pm$0.27 & $-$2.22$\pm$0.34  \\
		\hline
	\end{tabular}
\end{table*}

The best-fitting value of $U_\odot$ is consistent with some recent estimates, e.g., $12.00\pm0.60\,\rm km\,s^{-1}$ by Breddels et al. (2010) and $11.10_{-0.75}^{+0.69}\,\rm km\,s^{-1}$ by Sch{\"o}nrich et al. (2010). But this result is a slight larger than other recent estimates (Huang et al. 2015; Bobylev 2017; Bobylev \& Bajkova 2019; see our Table\,1). The value of $V_\odot$ is $10.16\pm0.51\,\rm km\,s^{-1}$, in great agreement with estimates of Bobylev \& Bajkova (2014) and Huang et al. (2015). We find $W_\odot=7.67\pm0.10\,\rm km\,s^{-1}$ with a small error bar, which is consistent with most of recent results shown in Table 1 (Breddels et al. 2010; Bobylev 2017; Bobylev \& Bajkova 2019).

Our results of non-zero Oort constants $C$ and $K$ suggest that the local Galactic disk is not symmetric. $|K+C|>4\,\rm km\,s^{-1}\,kpc^{-1}$ means that there is a radial velocity gradient in the extended local disk. The value of $C$, $-3.10\pm0.48\,\rm km\,s^{-1}\,kpc^{-1}$, is consistent with the results of $C=-3.2\pm0.4\,\rm km\,s^{-1}\,kpc^{-1}$ and $-2.7\pm0.1\,\rm km\,s^{-1}\,kpc^{-1}$ estimated by Bovy (2017) and Li et al. (2019), respectively. The value of $K$, $-1.25\pm1.04\,\rm km\,s^{-1}\,kpc^{-1}$, is slightly larger than $-3.3 \pm 0.6$\,km\,s$^{-1}$\,kpc$^{-1}$ derived by Bovy (2017), but in good agreement with $-1.7 \pm 0.2$\,km\,s$^{-1}$\,kpc$^{-1}$ given by Li et al. (2019). The measurements of $A$ and $B$ are also in agreement with $A=15.3\pm0.4\,\rm km\,s^{-1}\,kpc^{-1}$ and $B=-11.9\pm0.4\,\rm km\,s^{-1}\,kpc^{-1}$ (Bovy 2017). Our result of $B$ is larger than $B=-13.4\pm0.1\,\rm km\,s^{-1}\,kpc^{-1}$ estimated by Li et al. (2019). With Oort constants we can estimate the angular velocity $\Omega\,{\approx}\,A-B=28.30\pm1.19\,\rm km\,s^{-1}\,kpc^{-1}$. For $R_0=8.178\,\rm kpc$ (Gravity Collaboration et al. 2019), we can obtain the circular velocity  $V_{\rm c}=R_0\Omega=231.47\pm9.81\,\rm km\,s^{-1}$ and the solar rotation velocity $V_{\phi,\odot}=V_{\rm c}+V_\odot=241.63\pm9.82\,\rm km\,s^{-1}$. Our result of circular velocity is consistent with the values of $238\pm9\,\rm km\,s^{-1}$ and $240\pm6\,\rm km\,s^{-1}$ estimated by Sch{\"o}nrich (2012) and Huang et al (2016), respectively. The direct estimate of the solar angular velocity is the measurement of the proper motion of Sgr $\rm A^*$, $\Omega_\odot=30.24\pm0.12\,\rm km\,s^{-1}\,kpc^{-1}$ (Reid \& Brunthaler 2004). With the proper motion of Sgr $\rm A^*$, we can translate it to rotation velocity $V_{\phi,\odot}=247.30\pm1.44\,\rm km\,s^{-1}$ and our result is in agreement with this value. 

\begin{table}
	\centering
	\caption{The velocity dispersions for samples in different distance volumes.}
	\label{tab:example_table}
	\begin{tabular}{lccc} 
		\hline
		Range & $\sigma_{V_R}$ & $\sigma_{V_\phi}$ & $\sigma_{V_z}$\\
		$\,\rm (kpc)$ & $(\rm km\,s^{-1})$ & $(\rm km\,s^{-1})$ & $(\rm km\,s^{-1})$ \\
		\hline
		0-0.6 & 20.14$\pm$0.19 & 10.98$\pm$0.16 & 7.40$\pm$0.08\\
		0-0.7 & 19.95$\pm$0.17 & 10.90$\pm$0.13 & 7.45$\pm$0.06\\
		0-0.8 & 19.68$\pm$0.14 & 10.81$\pm$0.11 & 7.52$\pm$0.05\\
		0-0.9 & 19.51$\pm$0.11 & 10.59$\pm$0.10 & 7.50$\pm$0.05\\
		0-1.0 & 19.33$\pm$0.11 & 10.47$\pm$0.08 & 7.52$\pm$0.04\\
		\hline
	\end{tabular}
\end{table}

\begin{figure*}
	\centering
	\includegraphics[width = \linewidth]{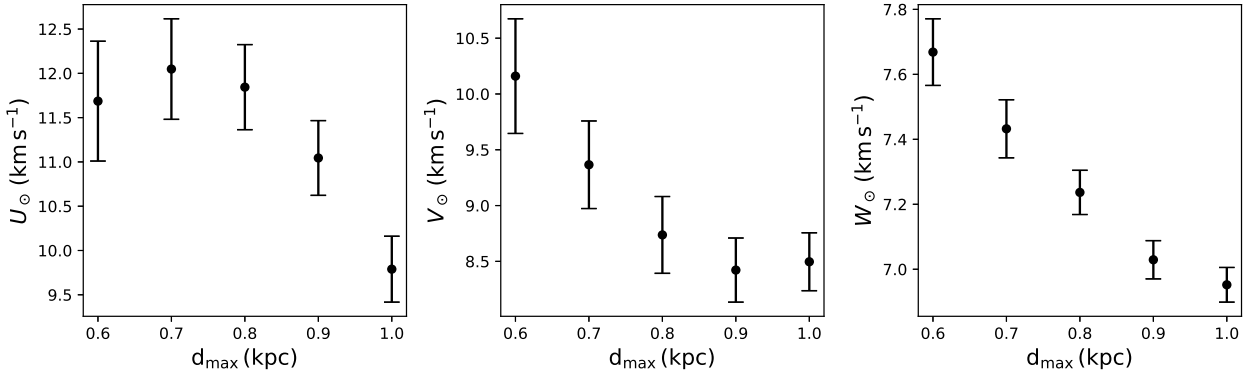}
	\caption{The results of solar peculiar velocities, $U_\odot$, $V_\odot$ and $W_\odot$, using the data in different volumes.  $d_{\rm max}$ represents the data within the range $(0-d_{\rm max})\,\rm kpc$.}
	\centering
	\label{fig:5}
\end{figure*}

\begin{figure*}
	\centering
	\includegraphics[width = \linewidth]{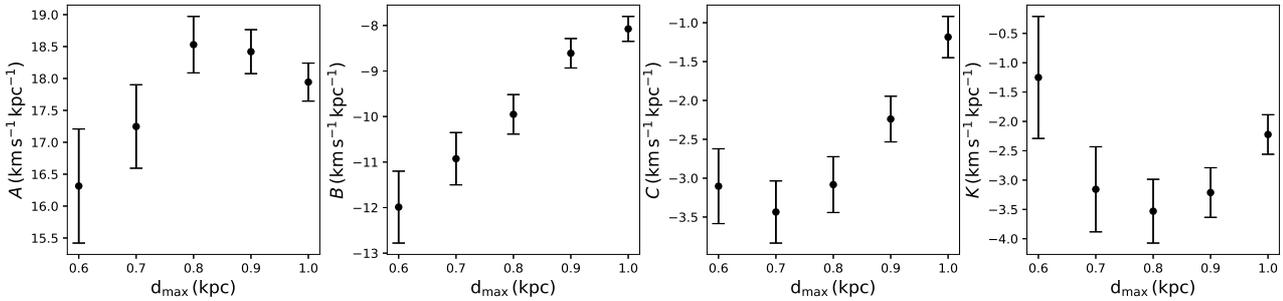}
	\caption{The results of Oort constants, $A$, $B$, $C$ and $K$, using the data in different volumes. $d_{\rm max}$ is same as that in Fig 5.}
	\centering
	\label{fig:6}
\end{figure*}
\section{Discussion}

Using a large sample with radial velocity, parallax and proper motions from Gaia DR2, Antoja et al. (2018) discovered clear ridge structures in the $R-V_\phi$ space and suggested that this phenomenon could be the signature of phase mixing. Wang et al. (2020) also recovered the ridge patterns in the $R-V_\phi$ plane color coded by mean $V_R$ and the ridge patterns were found from very young (OB stars, few hundred Myr) to very old populations ($9-14\,\rm Gyr$). These ridge structures may influence estimates of the solar motions and Oort constants. To evaluate the potential effects of the ridge structure on our derived parameters, we repeat the fitting analysis for our A-type star sample of different volumes which have different distances and the same vertical distance cut $|z|<0.1\,\rm kpc$. Here we only consider the ridge patterns toward the anti-center direction due to the spatial distribution of our sample. We define $d_{\rm max}$ to represent the volume of the adopted sample with $d < d_{\rm max}$. The results of the derived solar motions and Oort constants by different sample volumes (i.e., different $d_{\rm max}$) are presented in Table\,2 and Figs.\,5 and 6. Table\,3 shows the results of the velocity dispersion. There are clear trends between parameters, like $V_{\odot}$, $W_{\odot}$ and $B$, and $d_{\rm max}$ in Figs.\,5 and 6. Other parameters, i.e., $U_{\odot}$, $A$, $C$ and $K$ show weak variations along $d_{\rm max}$. In addition to above parameters, the determinations of local angular velocity (i.e., $A-B$) and the slope of local rotation curve ($\left. \frac{{\partial}V_{\rm c}}{{\partial}R}\right |_{R_0}=-A-B$) are shown in Fig.\,7. The former shows a weak variations along $d_{\rm max}$ while a clear decrease trend along $d_{\rm max}$ is detected for the later. The strong trend detected for the slope of the local rotation curve is expected given the ridge nature ($V_{\phi}$ decrease faster with $R$) found by Antoja et al. (2018) and Kawata et al. (2018).

According to the estimated parameters changing with $d_{\rm max}$, both the previous results and our result are probably affected by the external/internal perturbations on the Galactic disk, and the systematic deviations are around $0.5-3\,\rm km\,s^{-1}$ for $U_\odot$, $V_\odot$ and $W_\odot$ and $0.5-4\,\rm km\,s^{-1}\,kpc^{-1}$ for $A$, $B$, $C$ and $K$ for the current estimates, as revealed by Figs.\,5-7.

To try understand the systematic trends between estimated parameters and $d_{\rm max}$, we first made a very simple toy model to test the method adopted in the current work (see Section\,3 for details).
Generally, our model works quite well on estimating the solar peculiar velocities and Oort constants for stars within $1\,\rm kpc$ and certainly can not produce the variations of the estimated parameters changing with $d_{\rm max}$ as shown in Figs.\,5-7. The deviations of the estimated parameters are largely smaller than $0.5\,\rm km\,s^{-1}$ for $U_\odot$, $V_\odot$ and $W_\odot$ and $0.25\,\rm km\,s^{-1}\,kpc^{-1}$ for $A$, $B$, $C$ and $K$.
Moreover, the toy model shows that parameters derived from star sample within 0.6\,kpc are totally unbiased.

Secondly, we have preformed a mock data test to show such trends are probably caused by the external/internal perturbations on the Galactic disk.
The mock data is made by 1) an axisymmetric disk with $V_{z}$ warped as found recently (Schonrich et al. 2018; Huang et al. 2018; Li et al. 2020) and 2) a ridge pattern in $R$--$V_{\phi}$ plane (Antoja et al. 2018; Kawata et al. 2018; Khanna et al. 2019; Wang et al. 2020). We note the fraction of stars on the ridge pattern increases with $R$ (see Fig.\,1 of Kawata et al. 2018). More details about the mock data can be found in Appendix\,A.
We then apply our method to the mock data.
Interestingly, as shown in Figs.\,A1-3,  the trends between the estimated parameters and $d_{\rm max}$ are almost similar to the results found in Figs.\,5-7.

Our mock data test show that the perturbation signals found recently (i.e., the ridge pattern and the kinematic warp signal on $V_{z}$) can largely explain the trends between the estimated parameters (i.e., the peculiar velocities and  Oort constants) and $d_{\rm max}$.
However, we can not rule out other possibilities, e.g., the incomplete sky coverage of the current A-type star sample and other undiscovered perturbations on the Galactic disk.
We note further numerical simulations (with disk perturbation dynamics considered) are required to fully understand those trends (especially the larger variations found in observations than those in the current simulations) and accurately derive the peculiar velocities and Oort constants.

\begin{figure*}
	\centering
	\includegraphics[width = \linewidth]{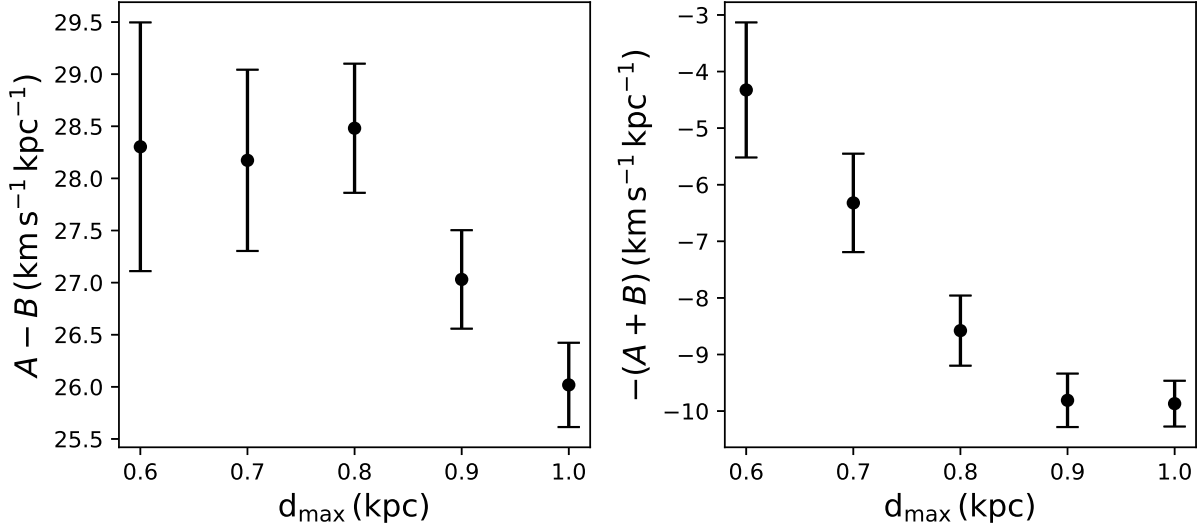}
	\caption{The results of $A-B$ and $-(A+B)$, which separately represent the angular velocity $\Omega$ and the slope of the circular velocity $\left. \frac{{\partial}V_{\rm c}}{{\partial}R}\right |_{R_0}$, using the data in different volumes. $d_{\rm max}$ is same as that in Fig.\,5.}
	\centering
	\label{fig:7}
\end{figure*}

\section{Conclusion}
In this work we derive the solar peculiar velocities and Oort constants by using a relatively young sample, A-type stars, selected from the value-add catalog of LAMOST DR4 and Gaia DR2. We adopt a MCMC method to fit 5,627 A-type stars within $d<0.6\,\rm kpc$ and $|z|<0.1\,\rm kpc$ using a non-axisymmetric model. We obtain the solar peculiar velocity $(U_\odot,V_\odot,W_\odot)=(11.69\pm0.68, 10.16\pm0.51, 7.67\pm0.10)\,\rm km\,s^{-1}$. The results of Oort constants are $A=16.31\pm0.89\,\rm km\,s^{-1}\,kpc^{-1}$, $B=-11.99\pm0.79\,\rm km\,s^{-1}\,kpc^{-1}$, $C=-3.10\pm0.48\,\rm km\,s^{-1}\,kpc^{-1}$, $K=-1.25\pm1.04\,\rm km\,s^{-1}\,kpc^{-1}$. $|K+C|>4\,\rm km\,s^{-1}\,kpc^{-1}$ means that there is a radial velocity gradient in the extended local disk, implying the local disk is in a non-asymmetric potential. The angular velocity $\Omega$ is $28.30\pm1.19\,\rm km\,s^{-1}\,kpc^{-1}$ estimated by $A-B$. For $R_0=8.178\,\rm kpc$, we can obtain the circular velocity $V_{\rm c}=231.47\pm9.81\,\rm km\,s^{-1}$ and the solar rotation velocity $V_{\phi,\odot}=V_{\rm c}+V_\odot=241.63\pm9.82\,\rm km\,s^{-1}$.

We also study the the effects of the ridge-like structures on constraining the solar peculiar velocities and Oort constants. We find that the ridge structures possibly result in larger values of $A$ and $B$, lower values of $V_\odot$ and a steeper slope of circular velocity as the volume becomes larger toward the anti-center direction and these effects are probably caused by ridge structure proportion increasing with distance. Detail dynamical model of the ridge structure is required to derive more accurate and unbiased solar peculiar velocities and Oort constants in the future.

\section*{Acknowledgements}
Thanks Dr. Xiaodian Chen for kind helps and discussions. This work is supported by the National Key R\&D Program of China No. 2019YFA0405500 and the National Natural Science Foundation of China under grant number 11973001. Y.H. is supported by National Natural Science Foundation of China grants 11903027, 11833006, 11811530289, and U1731108, and the Yunnan University grant C176220100007. H.F.W. is supported by the LAMOST Fellow project, China Postdoctoral Science Foundation via grant 2019M653504, Yunnan province postdoctoral Directed culture Foundation and the Cultivation Project for LAMOST Scientific Payoff and Research Achievement of CAMS-CAS.

Guoshoujing Telescope (the Large Sky Area Multi-Object Fiber Spectroscopic Telescope LAMOST) is a National Major Scientific Project built by the Chinese Academy of Sciences. Funding for the project has been provided by the National Development and Reform Commission. LAMOST is operated and managed by the National Astronomical Observatories, Chinese Academy of Sciences.

This work has made use of data from the European Space Agency (ESA) mission
{\it Gaia} (\url{https://www.cosmos.esa.int/gaia}), processed by the {\it Gaia}
Data Processing and Analysis Consortium (DPAC,
\url{https://www.cosmos.esa.int/web/gaia/dpac/consortium}). Funding for the DPAC
has been provided by national institutions, in particular the institutions
participating in the {\it Gaia} Multilateral Agreement.

\section*{DATA AVAILABILITY}
The data underlying this article are available in the LAMOST DR4 Value Added Catalogs, at \url{http://dr4.lamost.org/v2/doc/vac} and the ESA Gaia Archive, at \url{https://gea.esac.esa.int/archive}.

\newpage
\appendix
\section{Mock data test}
Here we make mock data in different volumes to explore the impacts of the ridge pattern (on $R$-$V_{\phi}$ plane) and the kinematic warp (on $V_{z}$) on estimating the solar peculiar velocities and the Oort constants. The mock data is made by 1) an axisymmetric disk with $V_{z}$ warped as found recently (Schonrich et al. 2018; Huang et al. 2018; Li et al. 2020) and 2) a ridge pattern on the $R$--$V_{\phi}$ plane (Antoja et al. 2018; Kawata et al. 2018; Khanna et al. 2019; Wang et al. 2020).
As shown in Fig.\,1 of Kawata et al. (2018),  the fraction of stars on the ridge pattern are assumed to increase with $R$.

For the mock data of the axisymmetric disk, we use the same values of $l$, $b$ and $d$ with our sample to keep the same sky coverage. As for the velocities, we utilise a simple model for the kinematics of Galactic disk. In the Galactocentric cylindrical coordinate, the mean values of $V_R$ and $V_z$ at different $R$ and $z$ are $0\,\rm km\,s^{-1}$. The rotation velocity at different $R$ is $V_{\phi}(R)=V_{\rm c}-3.4(R-R_\odot)\,\rm km\,s^{-1}$ which $V_{\rm c}=238\,\rm km\,s^{-1}$ is from Sch{\"o}nrich (2012) and ${\partial}V_{\rm c}/{\partial}R=-3.4\,\rm km\,s^{-1}\,kpc^{-1}$ is from Bovy (2017). We also adopt the $R_0=8.178\,\rm kpc$ (Gravity Collaboration et al. 2019).  The $V_R$, $V_z$ and $V_\phi$ of mock data are generated by the Gaussian random with the $\mathcal{N}(0,20)$,  $\mathcal{N}(V_{\phi}(R),10)$ and $\mathcal{N}(0,7)$ where 20, 10 and $7\,\rm km\,s^{-1}$ represent the velocity dispersions of the corresponding velocities.

For the ridge-like structure, the rotation velocity has a larger slope and the $V_R$ is nearly a constant (Wang et al. 2020), so we adopt $V_\phi=238-10(R-R_\odot)\,\rm km\,s^{-1}$ and $V_R=-4\,\rm km\,s^{-1}$ for the model. The velocity dispersions are $5$, $20$ and $7\,\rm km\,s^{-1}$ for $V_\phi$, $V_R$ and $V_z$, separately. The $(l,b,d)$ are randomly selected from our A-type star sample. Because some results (e.g. Fig.\,1 of Kawata et al. 2018) suggest that the ridge component becomes larger with distance in the anti-center direction within $1\,\rm kpc$, we consider the proportion of ridge-like structure increases with distance, e.g. the ratio of the ridge to axisymmetric disk data linearly increasing from 0.2:1 to 0.9:1 with the distance increasing from $0.6$ to $1.0\,\rm kpc$. Besides, for the kinematic signature of the Galactic warp which means the $V_z$ increasing with $R$ (Li et al. 2020), we just utilise a simple relation $V_z=1.0(R-R_\odot)\,\rm km\,s^{-1}$.

Then we transform the velocities of mock data to $V_r$, $V_l$ and $V_b$ with $(U_\odot,V_\odot,W_\odot)=(11.10, 12.24, 7.25)\,\rm km\,s^{-1}$ (Sch{\"o}nrich et al. 2010). The uncertainties of $V_r$ are randomly sampled from a uniform distribution from 5 to $7\,\rm km\,s^{-1}$ and the uncertainties of $V_l$ and $V_b$ are from 3 to $5\,\rm km\,s^{-1}$. Then we fit the mock data with our method. We repeat this process 100 times to avoid the random effects. 

The fitting results are shown in Figs.\,A1, A2 and A3. We find that the ridge proportion increasing with distance causes that $A$ and $B$ increase and $V_\odot$ and $-(A+B)$ decrease with $d_{\rm max}$. For $W_\odot$, the result has a clear decreasing trend. We find the influence of ridge structure on the $A-B$ is smaller than other parameters.

\begin{figure*}
	\centering
	\includegraphics[width = \linewidth]{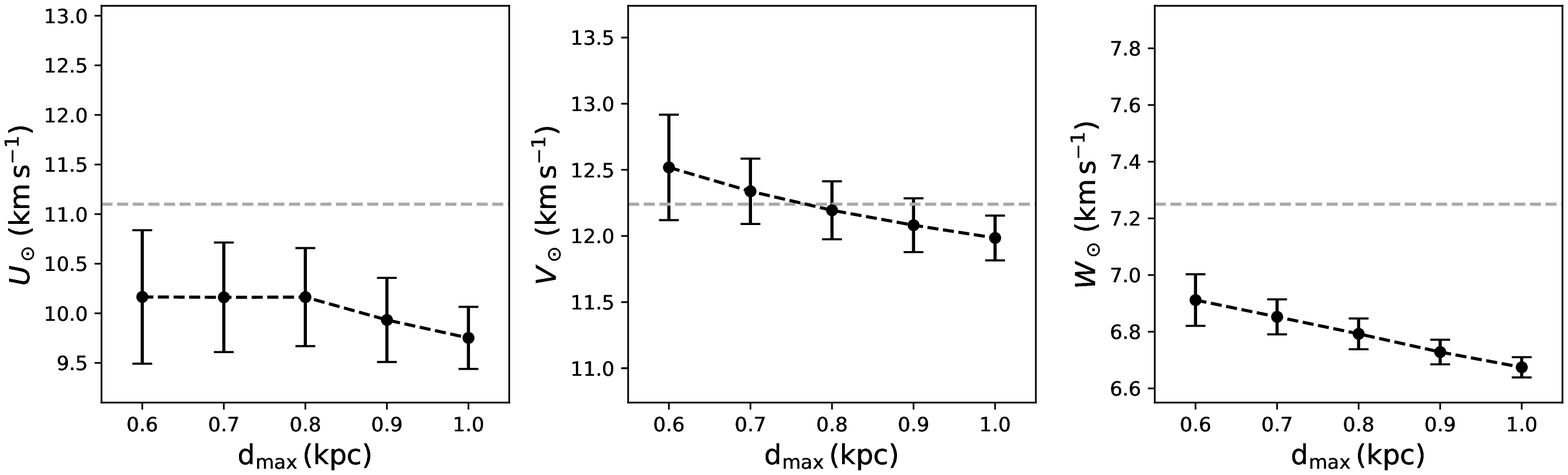}
	\caption{The results of solar peculiar velocities, $U_\odot$, $V_\odot$ and $W_\odot$, using the mock data in different volumes. $d_{\rm max}$ is same as that in Fig 5. The black dots and dashed lines represent our results of mock data in different volumes. The horizontal gray dashed lines represent the default values we use in the Galactic disk model.}
	\centering
	\label{fig:8}
\end{figure*}

\begin{figure*}
	\centering
	\includegraphics[width = \linewidth]{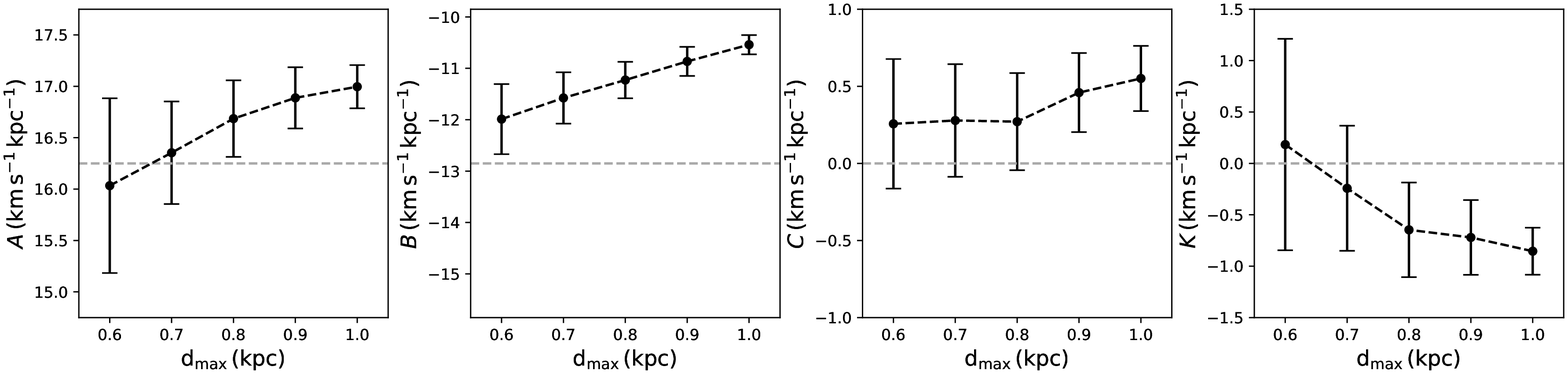}
	\caption{Same as Figure A1 but for the results of Oort constants, $A$, $B$, $C$ and $K$. The default values of $A$ and $B$ are calculated through $V_{\rm c}$, $R_0$ and ${\partial}V_{\rm c}/{\partial}R$.}
	\centering
	\label{fig:9}
\end{figure*}

\begin{figure*}
	\centering
	\includegraphics[width = \linewidth]{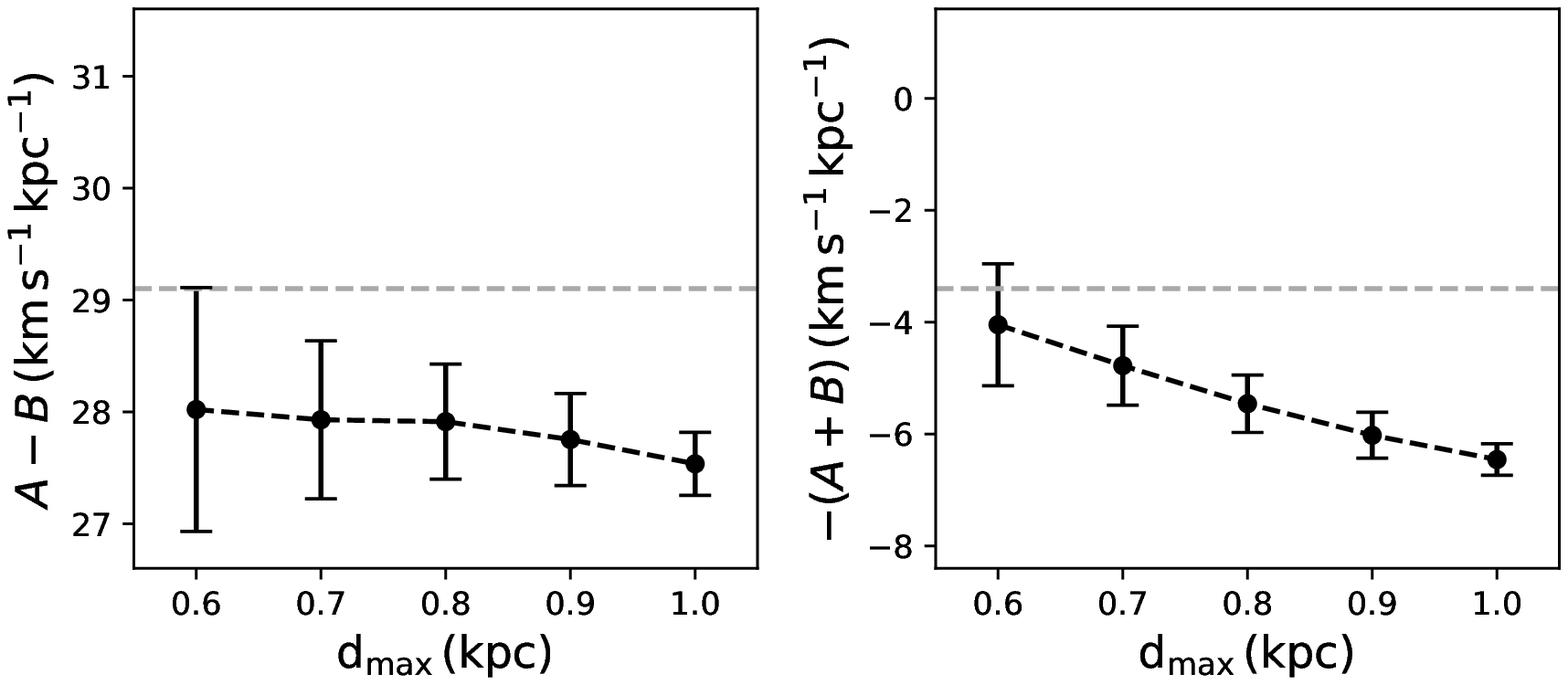}
	\caption{Same as Figure A1 but for the results of $A-B$ and $-(A+B)$ using the mock data in different volumes.}
	\centering
	\label{fig:10}
\end{figure*}

\end{document}